\documentclass[prl,showpacs,twocolumn,showkeys,amssymb,amsmath,aps]{revtex4}

\usepackage{graphicx}
\usepackage{bm}
\begin{document}
\title{Generalized detailed Fluctuation Theorem under Nonequilibrium  
        Feedback control}
\author{M. Ponmurugan\footnote{email correspondence: mpn@imsc.res.in}} 
\affiliation{The Institute of Mathematical Sciences, 
C.I.T. Campus, Taramani, Chennai 600113, India}
\date{\today}

\begin{abstract} 
It has been shown recently that the Jarzynski equality 
is generalized  under nonequilibrium feedback control 
[T. Sagawa and M. Ueda, Phys. Rev. Lett. {\bf 104}, 090602 (2010)].
The presence of feedback control in physical systems 
should modify both Jarzynski equality and detailed fluctuation theorem
[K. H. Kim and H. Qian, Phys. Rev. E {\bf 75}, 022102 (2007)].
However, the generalized Jarzynski equality under forward feedback control
has been proved by consider that the physical systems under 
feedback control should locally satisfies
the detailed fluctuation theorem.  
We use the same formalism and derive the generalized 
detailed fluctuation theorem under nonequilibrium feedback control.  
It is well known that the exponential average 
in one direction limits the calculation of precise free 
energy differences. The knowledge of measurements from  
both directions usually gives improved results. 
In this aspect, the generalized detailed fluctuation theorem
can be very useful in free energy calculations  for 
system driven under nonequilibrium feedback control.  
 
\end{abstract}
\maketitle

Recent development in statistical physics relationships
describe nonequilibrium dynamics in terms of 
equalities \cite{jar,crooks,other}.
In particular, the  Jarzynski's equality \cite{jar} and the 
Crooks detailed fluctuation theorem \cite{crooks} can be used
to calculate equilibrium free energy differences from
nonequilibrium work distribution. A system initially
 at equilibrium with temperature (inverse) $\beta=1/k_BT$
($k_B$ is the Boltzmann constant) is externally driven from 
its initial state to final state by nonequilibrum 
process satisfies the detailed fluctuation theorem 
$P(W)/P(-W) = \mbox{exp}(\beta (W- \Delta F))$ and its integrated version,
the Jarzynski equality $\langle \mbox{exp}(-\beta W) \rangle =\mbox{exp}(-\beta \Delta F)$.
Where $W$ denotes the work performed on the system, $\Delta F$ 
is the free energy  difference of the system between its final 
and initial equilibrium states and $P(\pm W)$ is the work probability 
distribution in forward (+) and reverse (-) direction.
The (exponential) average $\langle.\rangle$ is taken over the ensemble of
nonequilibrium trajectories. 
These relationships has been verified in experiments \cite{lip,col}
as well as simulations \cite{steer,Janosi} and widely used in 
many branches of Science (see, eg.\cite{van}).

Various experiments and simulations has been performed by
adopting a suitable time-dependent driving scheme 
described by an external control switching  protocol.
Even though the Jarzynski equality and the detailed fluctuation 
theorem are valid for any time-dependent driving scheme, 
the efficiency of a nonequilibrium switching simulations 
which use these relationships to estimate precise free energy difference
is depends on the switching function \cite{seif}.
The practical difficulties faced in precise 
calculation of free energy differences in simulations \cite{fdiff}
initiated further developments in these theories \cite{jar1}. 
In our recent work, we have discussed the analogy between the
optimized switching free energy simulation  
and the nonequilibrium systems which are subjected to 
a feedback control \cite{pon}.

The evolution of the physical systems can be modified or controlled 
by repeated operation of an external agent called controller \cite{feedref,frev}.
In contrast to open loop controller which operates on the 
system blindly, the feedback or closed loop controllers use 
information about the state of the system. The feedback
is the process performed by the controller of measuring the 
system, deciding on the action given the measurement output, 
and acting on the system \cite{cao}. For example, in a single molecule
Atomic Force Microscopy experiment, the external agent 
is an electric feedback circuit which detects the motion of  
the cantilever and manipulate 
the control force proportional to its velocity \cite{qian}. 
The proper utilization of the information about the 
state of the system in feedback control effectively 
improves the system performance \cite{feedref,frev,cao,qian,sagab1}. 
However, the presence of feedback control in physical system modifies both the 
Jarzynski equality and the fluctuation theorem \cite{qian}.

Recently, the Jarzynski equality is generalized 
to an experimental condition in which the system is driven between 
two equilibrium state via nonequilibrium process  
under forward feedback control \cite{sagaf}.
Since the work that performed on a thermodynamic system 
can be lowered by feedback control \cite{sagaf}, this  
feedback mechanism  can be helpful in simulation for 
sampling rare trajectories and calculate  
precise free energy differences \cite{pon}. 
The equilibrium free energy difference for the driven  system (which 
locally satisfies the detailed fluctuation theorem) under nonequilibrium 
feedback control in forward direction can be calculated from the 
generalized Jarzynski equality \cite{sagaf}
\begin{eqnarray}\label {equl1}
\langle e^{- \sigma - I} \rangle &=& 1, 
\end{eqnarray}
where $\sigma = \beta (W - \Delta F)$ and $I$ 
is the mutual information measure obtained by the 
feedback controller \cite{sagaf}. The average is 
taken from the work distribution  in forward 
direction  with feedback control.

The second law of thermodynamics can be quantitatively
described by the fluctuation theorem which are closely 
related to the Jarzynski equality. If the experiments or simulations
has been performed under feedback control, both the Jarzynski equality
and the fluctuation theorem should be extended \cite{qian}. 
However, the generalized Jarzynski equality (Eq.\ref{equl1})
under forward feedback control has been proved by consider that 
the physical systems under 
feedback control should locally satisfies
the detailed fluctuation theorem \cite{sagaf}. 
In this letter, we use the same formalism and derive the 
generalized detailed fluctuation theorem under nonequilibrium 
feedback control.
In order to calculate the free energy differences precisely 
in simulations one require information of work 
distribution in both forward and reverse direction 
\cite{compare1,compare2,zucker}. In this aspect,
one needs the generalized detailed fluctutaion 
thoerem under nonequilibrium feedback control.

The feedback control enhances our controllability of 
small thermodynamics systems \cite{feedref,frev,cao,qian,sagab1}. 
At a given time, the controller measure the 
partial state of the system. The result of the measurement 
determines the action the control will take. 
The additional information on the system 
provided by the measurement further 
determines the system states.
Suppose, the controller perform a measurements on a  stochastic 
thermodynamics system at time $t_m$. Let $\Gamma_m$ be the 
phase-space point of the system at that time,
$P[\Gamma_m]$ its probability and $y$ the measurement.
Depends on the controller measurements  
the measurement outcome $y$ occurs 
with probability $P[y]$ \cite{sagaf}.
The information obtained by the controller measurement can be characterized
by the mutual (feedback) information measure \cite{sagaf,beck},
\begin{eqnarray}\label{iid}
I[y,\Gamma_m]=ln \left[ \frac{P[y|\Gamma_m]}{P[y]} \right ],
\end{eqnarray}
where $P[y|\Gamma_m]$ is the conditional probability of 
obtaining outcome $y$ on condition that the state of the system 
is $\Gamma_m$. 
 The above equation is rewritten as, 
\begin{eqnarray}\label{eid}
e^{I[y,\Gamma_m]} &=& \frac{P[y|\Gamma_m]}{P[y]}.
\end{eqnarray}

In experiments and simulations, the free energy difference 
between the two equilibrium states can be calculated 
in general by pulling the sytem from one equilibrium state 
to another state along a switching path. 
The path connecting the two states in the 
time period $\tau$  will be parameterized  
using the variable $\lambda$, with $0 \le \lambda \le 1$.
The switching rate describes the nature of the switching 
process to be an equilibrium (infinitely slow) or nonequilibrium (fast).
If the experiments performed under feedback control,
the switching control parameter $\lambda$ depends on the outcome 
$y$ after $t_m$ \cite{sagaf}.
That is, whenever the controller made measurements, there 
is a corresponding changes in the switching parameter
for next time step, which is denoted as $\lambda_{(t;y)}$.
Between the every stages of the controller measurements, 
the value of outcome $y$ is fixed and the corresponding 
switching paremeter $\lambda_{(t;y)}$ doesnot change. 
In this time interval for each stages  of $\lambda_{(t;y)}$,
we consider the system  should locally satisfies the detailed 
fluctuation theorem \cite{sagaf},
\begin{eqnarray}\label{locfluc}
\frac{P_{\lambda_{(t;y)}}[\Gamma(t)]}{P_{\lambda_{(t;y)}^{\dagger}}[\Gamma^{\dagger}(t)]} 
&=& e^{\sigma[\Gamma(t)]},
\end{eqnarray}
where $P_{\lambda_{(t;y)}}[\Gamma(t)]$ is the 
probability of obtaining the outcome $y$ in forward direction 
with switching protocol $\lambda_{(t;y)}$ and  
$P_{\lambda_{(t;y)}^{\dagger}}[\Gamma^{\dagger}(t)]$
is the probability of obtaining the same outcome $y$ \cite{sagaf}
in reversed direction of phase point, $\Gamma^{\dagger}(t)$,  
with corresponding switching protocol $\lambda^{\dagger}_{(t;y)}$.
Here, $\sigma[\Gamma(t)]$ is the work value obtained in the 
forward direction and its time reversal work value 
\begin{eqnarray}\label{worksym}
\sigma[\Gamma^{\dagger}(t)]&=& - \sigma[\Gamma(t)].
\end{eqnarray}

Let $P_F[\widetilde{X}] \equiv P_F[\sigma[\widetilde{\Gamma}],I[\widetilde{y},\widetilde{\Gamma}]]$ be the 
the joint probability of obtaining the work value $\sigma[\widetilde{\Gamma}]$ for a given 
feedback information measure $I[\widetilde{y},\widetilde{\Gamma}]$
of measurement outcome $\widetilde{y}$ in forward direction.
Due to the repeated measurements of the controller, the
particular measurement outcome $\widetilde{y}$ may occurs in various stages of 
the experiment. 
In analogous to previous studies \cite{jarq,adibq}, 
this (joint) probability can be obtained from the 
nonequilibrium ensemble of  variable (forward) switching trajectories as, 
\begin{eqnarray}\label{fl1}
P_F[\widetilde{X}] &=& \int  P[y'|\Gamma'_m] P_{\lambda_{(t;y')}}[\Gamma(t)] 
 \delta(I[y',\Gamma'_m]-I[\widetilde{y},\widetilde{\Gamma}]) \nonumber \\
 & & \delta(\sigma[\Gamma(t)]-\sigma[\widetilde{\Gamma}]) \ dy' \ {\it D}[\Gamma(t)],
\end{eqnarray}
where $\delta(x)$ is the Dirac delta function which has a property
$\delta(-x)=\delta(x)$. It should be noted that $y'$ and
$\Gamma'_m$ in the conditional probabilities are dummy variables
and it has appropriate values for each outcome of the 
controller measurements, see Eq.(\ref{eid}).

Combining Eq.(\ref{eid}) and  Eq.(\ref{locfluc}), then  Eq.(\ref{fl1}) 
becomes,
\begin{eqnarray}
P_F[\widetilde{X}] &=& \int
e^{\sigma[\Gamma(t)]+I[y',\Gamma'_m]}  P[y'] P_{\lambda_{(t;y')}^{\dagger}}[\Gamma^{\dagger}(t)] \nonumber \\
 & &\delta(I[y',\Gamma'_m]-I[\widetilde{y},\widetilde{\Gamma}]) 
 \delta(\sigma[\Gamma(t)]-\sigma[\widetilde{\Gamma}]) \nonumber \\
 & & dy' \ {\it D}[\Gamma(t)], \nonumber
\end{eqnarray}
\begin{eqnarray}\label{fl2}
P_F[\widetilde{X}] &=&
e^{\sigma[\widetilde{\Gamma}]+I[\widetilde{y},\widetilde{\Gamma}]} 
\int P[y'] P_{\lambda_{(t;y')}^{\dagger}}[\Gamma^{\dagger}(t)] \nonumber \\
 & &\delta(I[y',\Gamma'_m]-I[\widetilde{y},\widetilde{\Gamma}]) 
 \delta(\sigma[\Gamma(t)]-\sigma[\widetilde{\Gamma}]) \nonumber \\
 & & dy' \ {\it D}[\Gamma(t)].
\end{eqnarray}

Let $P_R[\widetilde{X}^{\dagger}] \equiv P_R[-\sigma[\widetilde{\Gamma}],I[\widetilde{y}^{\dagger},\widetilde{\Gamma}]]$ be the 
the joint probability of obtaining the work value $-\sigma[\widetilde{\Gamma}]$ for a given 
feedback information measure $I[\widetilde{y}^{\dagger},\widetilde{\Gamma}]$ of 
measurement outcome $\widetilde{y}^{\dagger}$
in reverse direction.
This (joint) probability can be obtained from the 
nonequilibrium ensemble of  variable (reverse) switching trajectories as, 
\begin{eqnarray}\label{fl3}
P_R[\widetilde{X}^{\dagger}] 
&=& \int  P[y'|\Gamma'_m] P_{\lambda_{(t;y')}^{\dagger}}[\Gamma^{\dagger}(t)] \nonumber \\
 & &\delta(I[y',\Gamma'_m]-I[\widetilde{y}^{\dagger},\widetilde{\Gamma}]) 
 \delta(\sigma[\Gamma^{\dagger}(t)]+\sigma[\widetilde{\Gamma}]) \nonumber \\
 & & dy' \ {\it D}[\Gamma^{\dagger}(t)].
\end{eqnarray}
To this end  we will use above equations and 
derive the generalized detailed fluctuation theorem under 
feedback control either in forward direction \cite{sagaf} 
or both directions as follows.

If the system has same feedback control in both directions
then under the controller measurements condition 
$I[\widetilde{y}^{\dagger},\widetilde{\Gamma}]=I[\widetilde{y},\widetilde{\Gamma}]$
and using Eq.(\ref{eid}), we can rewrite Eq.(\ref{fl3}) as,
\begin{eqnarray}\label{both1}
P_R[\widetilde{X}^{\dagger}] 
&=& e^{I[\widetilde{y},\widetilde{\Gamma}]} 
\int  P[y'] P_{\lambda_{(t;y')}^{\dagger}}[\Gamma^{\dagger}(t)] \nonumber \\
 & &\delta(I[y',\Gamma'_m]-I[\widetilde{y},\widetilde{\Gamma}]) 
 \delta(\sigma[\Gamma^{\dagger}(t)]+\sigma[\widetilde{\Gamma}]) \nonumber \\
 & & dy' \ {\it D}[\Gamma^{\dagger}(t)].
\end{eqnarray}
Since $\delta(-x)=\delta(x)$ and ${\it D}[\Gamma^{\dagger}(t)]={\it D}[\Gamma(t)]$ \cite{sagaf},
combining Eq.(\ref{worksym}) and Eq.(\ref{both1}) in Eq.(\ref{fl2}) 
we can obtain the generalized detailed fluctuation theorem
under feedback control in both direction as
\begin{eqnarray}
P_F[\widetilde{X}] &=& 
e^{\sigma[\widetilde{\Gamma}]+I[\widetilde{y},\widetilde{\Gamma}]} 
P_R[\widetilde{X}^{\dagger}]e^{-I[\widetilde{y},\widetilde{\Gamma}]}, \nonumber 
\end{eqnarray}
\begin{eqnarray}\label{genfluc2}
\frac{P_F[\widetilde{X}]}{P_R[\widetilde{X}^{\dagger}]}
 &=& e^{\sigma[\widetilde{\Gamma}]}. 
\end{eqnarray}
The above equation can be written simply as 
\begin{eqnarray}
\frac{P_F[\sigma, I]}{P_R[-\sigma, I]}
 &=& e^{\sigma}. 
\end{eqnarray}
Our result shows that for a given feedback information 
measure $I$ in both directions, 
the physical  system under feedback control satisfies the 
detailed fluctuation theorem
$\frac{P_F[\sigma]}{P_R[-\sigma]}=e^{\sigma}$. 
Since the feedback control enhances our controllability of 
small thermodynamics systems, the proper choice of 
feedback mechanism in free energy simulations 
can be useful for precise free energy estimates 
instead looking for optimized 
switching protocols \cite{pon} . 

If the system has feedback control only 
in forward direction, the generalized detailed 
fluctuation theorem under forward feedback control can be 
obtained from the following reverse experimental condition. 
Based on the informations 
about switching protocols for each outcome in  
forward feedback control experiment, 
without feedback control, we can perform same 
variable switching protocols 
experiment in reverse direction  \cite{sagaf}.
This is equivalent as an open loop controller 
which operates on the system in reverse direction.
In such a case,  $P[y'|\Gamma'_m]=P[y']$ in 
reverse direction
and the open loop controller implicitly 
has information about the forward 
feedback information measure 
$I[\widetilde{y},\widetilde{\Gamma}]$   
for corresponding  switching parameter, $\lambda_{(t;y')}^{\dagger}$,  
in reverse direction, see Eqs.(\ref{eid}) and (\ref{locfluc}). 
Then, we can write Eq.(\ref{fl3}) as 
\begin{eqnarray}\label{re1}
P_R[\widetilde{X}^{\dagger}] 
&=& \int  P[y'] P_{\lambda_{(t;y')}^{\dagger}}[\Gamma^{\dagger}(t)] \nonumber \\
 & &\delta(I[y',\Gamma'_m]-I[\widetilde{y},\widetilde{\Gamma}]) 
 \delta(\sigma[\Gamma^{\dagger}(t)]+\sigma[\widetilde{\Gamma}]) \nonumber \\
 & & dy' \ {\it D}[\Gamma^{\dagger}(t)].
\end{eqnarray}
As similar to earlier derivation, combining Eq.(\ref{worksym}) 
and Eq.(\ref{re1}) in Eq.(\ref{fl2}) 
we can obtain the generalized detailed fluctuation theorem
under forward feedback control as
\begin{eqnarray}
P_F[\widetilde{X}] &=& 
e^{\sigma[\widetilde{\Gamma}]+I[\widetilde{y},\widetilde{\Gamma}]} 
P_R[\widetilde{X}^{\dagger}], \nonumber 
\end{eqnarray}
\begin{eqnarray}\label{genfluc}
\frac{P_F[\widetilde{X}]}{P_R[\widetilde{X}^{\dagger}]}
 &=& e^{\sigma[\widetilde{\Gamma}]+I[\widetilde{y},\widetilde{\Gamma}]}. 
\end{eqnarray}

In order to prove the generalized Jarzynski equality
under forward feedback control,
we measure the quantity,
\begin{eqnarray}
\langle e^{-\sigma-I} \rangle &=&
\int P_F[\widetilde{X}] \ e^{-\sigma[\widetilde{\Gamma}]-I[\widetilde{y},\widetilde{\Gamma}]} 
\ d\widetilde{y} \ d\widetilde{\Gamma}.  \nonumber 
\end{eqnarray}
From Eq.(\ref{genfluc}),
\begin{eqnarray}
\langle e^{-\sigma-I} \rangle 
&=&  \int P_R[\widetilde{X}^{\dagger}]
 \ d\widetilde{y} \ d\widetilde{\Gamma} \nonumber \\
 &=& 1, 
\end{eqnarray}
we obtained the generalized Jarzynski equality 
under forward feedback control \cite{sagaf}.
In order to get  more insight of 
forward mutual information measure,
we calculate the  quantity \cite{sagaf} 
\begin{eqnarray}
\langle e^{- \sigma } \rangle &=& 
\int P_F[\widetilde{X}] \ e^{-\sigma[\widetilde{\Gamma}]} 
\ d\widetilde{y} \ d\widetilde{\Gamma}  \nonumber 
\end{eqnarray}
 Using Eq.(\ref{genfluc}), 
the average of above equation can be written as 
\begin{eqnarray}
\langle e^{- \sigma } \rangle &=& 
 \int P_R[\widetilde{X}^{\dagger}] \ e^{I[\widetilde{y},\widetilde{\Gamma}]} 
\ d\widetilde{y} \ d\widetilde{\Gamma}  \nonumber \\
&=& \int P_R[\widetilde{X^{\star}}^{\dagger}] 
\ d\widetilde{y} \ d\widetilde{\Gamma},  \nonumber \\
&=& \gamma, 
\end{eqnarray}
where $ \gamma =\int P_R[\widetilde{X^{\star}}^{\dagger}] \ d\widetilde{y} \ d\widetilde{\Gamma}$ 
is the feedback control characteristics which is 
a measure of the correlation between the dissipation 
and the information \cite{sagaf}
and $P_R[\widetilde{X^{\star}}^{\dagger}]=P_R[\widetilde{X}^{\dagger}]
e^{I[\widetilde{y},\widetilde{\Gamma}]}$
is the special case \cite{sagaf} of the joint probability 
distribution in reverse direction.  

Even though there is no feedback control in reverse direction,
due to the implementation of variable (reverse) switching protocols
in accordance with forward feedback control experiment, 
the forward mutual information measure can also obtained 
from the reverse direction as 
\begin{eqnarray}
I[\widetilde{y},\widetilde{\Gamma}]
 &=& ln \left[ \frac { P_R[\widetilde{X^{\star}}^{\dagger}]}
{P_R[\widetilde{X}^{\dagger}]} \right] .   
\end{eqnarray}

In conclusion, we have derived the generalized detailed 
fluctuation theorem under nonequilibrium feedback control.
It is well known that the exponential average 
in one direction limits the accurate calculation of free 
energy differences in simulation. The knowledge of measurements from  
both directions usually gives improved results. 
Thus, the generalized detailed fluctuation theorem
can be very useful in free energy simulation for 
system driven under nonequilibrium feedback control.

\end{document}